\documentclass[prd, amsfonts, twocolumn, nofootinbib, showpacs]{revtex4-1}
\usepackage{graphicx, epsfig}
\usepackage{color}
\usepackage{amsmath}
\usepackage{mathtools}
\newcommand{\be}{\begin{equation}}
\newcommand{\ee}{\end{equation}}
\newcommand{\bea}{\begin{eqnarray}}
\newcommand{\eea}{\end{eqnarray}}

\newcommand{\gapp}{\mathrel{\raise.3ex\hbox{$>$}\mkern-14mu \lower0.6ex\hbox{$\sim$}}}
\newcommand{\lapp}{\mathrel{\raise.3ex\hbox{$<$}\mkern-14mu \lower0.6ex\hbox{$\sim$}}}

\newcommand{\GMLfull}{Gell-Mann-L\'evy~}

\newcommand{\GMLmth}{{\mathrm GML}}

\newcommand{\HVEV}{\langle H\rangle}
\newcommand{\SVEV}{\langle S\rangle}
\newcommand{\half}{\frac{1}{2}}
\newcommand{\quarter}{\frac{1}{4}}
\newcommand{\mpisq}{m_\pi^2}
\newcommand{\mthsq}{m_3^2}
\newcommand{\mhsq}{m_h^2}
\newcommand{\hbbig}{\underbar h}
\newcommand{\hb}{\small \underbar h}
\newcommand{\mhbsq}{m_{\small \underbar h}^2}

\def\bbox{{\,\lower0.9pt\vbox{\hrule \hbox{\vrule height 0.2 cm
\hskip 0.2 cm \vrule  height 0.2 cm}\hrule}\,}}

\begin{document} 
\title{The Goldstone theorem protects naturalness, and the
 absence of \\ Brout-Englert-Higgs fine-tuning, in spontaneously broken SO(2)}
 \author{Bryan W. Lynn$^{1}$ and Glenn D. Starkman$^{2,3}$ }
\affiliation{$^1$ University College London, London WC1E 6BT, UK}
\affiliation{$^2$ ISO/CERCA/Department of Physics, Case Western Reserve University, Cleveland, OH 44106-7079}
\affiliation{$^3$ CERN, CH-1211 Geneva 23, Switzerland}
\email{bryan.lynn@cern.ch, gds6@case.edu}
%

\begin{abstract}

The \GMLfull\!\! (GML), Schwinger and Standard Models
were previously shown not to suffer from a Brout-Englert-Higgs (BEH) fine-tuning problem due to ultraviolet quadratic divergences, with finite Euclidean cut-off $\Lambda$, because of the symmetries obeyed by all ${\cal O}(\Lambda^2)$ contributions.
We extend those no-fine-tuning results to finite contributions from certain $M_{\rm Heavy}^2\gg m_{\rm BEH}^2$ particles, in simplified SO(2) versions of GML and Schwinger.
We demonstrate explicit 1-loop physical naturalness (i.e. G. 't Hooft's 1979 criteria for no-fine-tuning) for two examples, both SO(2) singlets:
a heavy $M_S^2\gg m_{\rm BEH}^2$ real scalar $S$ with discrete $S\to -S$ symmetry;
a right-handed Type 1 See-saw Majorana neutrino $\nu_R$ with $M_{\nu_R}^2\gg m_{\rm BEH}^2$. 
We prove that for  $\left|q^2\right| \ll M_{Heavy}^2$
the heavy degrees of freedom contribute, at worst, marginal operators in spontaneously broken SO(2) Schwinger.

The key \GMLfull lesson, from these two one-loop examples, is that the pseudo Nambu-Goldstone boson (NGB) mass-squared $\mthsq$ must be properly renormalized. A true NGB value, $\mthsq = 0$, is then protected by the Goldstone theorem.
For the SO(2) Schwinger model, two crucial observations emerge: that global Ward-Takahashi identities (WTI) force all relevant operators into Schwinger's symmetric Wigner mode, i.e. into the pseudo-NGB mass-squared $\mthsq \neq 0$; and WTI
enforce the Goldstone theorem by forbiding all relevant operator contributions -- ${\cal O}(\Lambda^2)$, ${\cal O}(M_{Heavy}^2 \ln \Lambda^2)$,${\cal O}(M_{Heavy}^2)$ and ${\cal O}(M_{Heavy}^2 \ln M^2_{Heavy})$ -- 
in Schwinger's spontaneously broken Goldstone mode,
i.e. $\pi_3$ is a true NGB there with $\mthsq = 0$. GML/Schwinger Goldstone mode, with weak-scale $m_{BEH}^2,\HVEV^2 \ll M_{Heavy}^2$, is not-fine-tuned, not just as a stand-alone renormalizable field theory,
but also as a low-energy effective theory with certain high-mass-scale extensions.
Its "Goldstone Exceptional Naturalness", where all relevant operators vanish identically, is a far more powerful suppession of fine-tuning (i.e. than G 't Hooft's criteria) and is simply another (albeit un-familiar) consequence of WTI, spontaneous symmetry breaking and the Goldstone theorem. 

If Goldstone Exceptional Naturalness can somehow be extended to the Standard Model (SM), 
there should be no expectation that LHC will discover any Beyond the SM physics
unrelated to neutrino mixing, i.e. the only known experimentally necessary modification of the Standard Model plus 
General Relativity paradigm.

\end{abstract}

\maketitle


Naturalness has played a central role in theoretical particle physics over the last several decades.
Theories deemed ``un-natural'' are condemned as fatally Fine-Tuned (FT).  
It was the widespread
belief that the scalar sector of the Standard Model (SM) suffered from  the so-called
Brout-Englert-Higgs (BEH) FT problem \cite{Susskind1979,Wilson} that drove the development of many  Beyond the Standard Model (BSM)
theories, including Low Energy Supersymmetry and Technicolor.     
This letter takes a step (i.e. for global theories) toward a non-BSM proposal, based on the Goldstone theorem, that may resolve a perceived crisis \cite{Lykken2104} due to tension between LHC8 data and simple BSM solutions of that ``BEH-FT problem."

One challenge has been pinning down exactly what it means to be ``natural,'' 
what it means to be ``fine-tuned,'' and how they are related.
In this letter we will demonstrate how, when fine-tuned and natural are regarded as mutually exclusive, 
not only are the \GMLfull Model (GML) \cite{GellMannLevy1960}  and the Schwinger model \cite{Schwinger1957} 
not FT \cite{Lynnetal2012} with respect to ultraviolet quadratic divergences (UVQD), but neither are a variety of interesting extensions to GML and Schwinger. 
The key is for any new high-mass-scale physics to respect the same symmetry that protects GML and Schwinger.  

We call G. 't Hooft's widely accepted naturalness/no-fine-tuning criteria {\bf physical naturalness} (PhysNat) because it avoids all reference to unmeasurable quantities: ``at any energy scale $\mu$ a [dimensionless] physical parameter or a set of physical parameters $\alpha_i(\mu)$ 
is allowed to be very small only if the replacement $\alpha_i(\mu)=0$
would  increase the symmetry of the system"  \cite{tHooft1980, AlvarezGaume2012}.

We define here a new, far more powerful, suppression of fine-tuning called {\bf Goldstone Exceptional Naturalness} (GoldExceptNat): ``If taking a dimensionless parameter to zero increases the symmetry of the system, and that increased symmetry then forbids all relevant-operator contributions to observables or counter-terms, then a small value for that quantity, and the resultant theory, are GoldExceptNat. GoldExceptNat is simply another (albeit unfamiliar) consequence of Ward-Takahashi identities (WTI), spontaneous symmetry breaking (SSB) and the Goldstone therorem." 

Theories are ordinarily regarded as ``Bare Fine Tuned" if physical observables depend too sensitively on
the parameters of the bare Lagrangian $L^{bare}$, especially if small dimensionless ratios
of physical observables arise through cancellation of large (often divergent, and generally including bare) contributions 
unrelated by any symmetry.  
Bare-FT troubles us.
$L^{bare}$ is only a useful parametric device,
a calculation tool without physical reality of its own, expressed in terms of unphysical bare parameters, whose
values cannot be determined by any set of experiments.
To judge a theory based on how a mathematical fiction depends on unphysical parameters seems unjustified.
We argue here, both for the stronger PhysNat definition
that avoids reference to unmeasurable quantities,  and to explicitly exclude natural theories from being FT.  
If taking a dimensionless quantity to zero increases the symmetry of a theory, 
then a small value for that quantity is natural and not  FT.

Some time ago, we demonstrated that the SM \cite{Lynn2011}
and its naive zero-gauge coupling limit in the scalar sector, 
the chiral-symmetric limit of GML \cite{Lynnetal2012}, are UVQD-PhysNat (and consequently are  not FT). 
This is because in GML all cancellations of UVQD, to obtain weak-scale quantities,
are absorbed into the mass-squared $\mpisq$ (at $q^2=0$) of the three  {\bf pseudo} Nambu-Goldstone bosons (pseudo-NGBs) that share the same SU(2) doublet as the BEH.   
Other dimensionful contributions to physical quantities are proportional to $\HVEV^2$, the square of the vacuum expectation value (VEV) of the BEH field,
  and to dimensionless couplings that can be fixed by low-energy experiments.
Since taking {\bf either} $\mpisq$ or $\HVEV^2$ to zero restores chiral $SU(2)_{L-R}$ symmetry,
the smallness of $\mpisq$ or $\HVEV^2$ compared to a Euclidean-integral UV cutoff-squared $\Lambda^2$ is PhysNat,
not FT.   (Note that UVQD are the only potential source of large dimensionful quantities in the SM \cite{Lynn2011}, 
since the SM has no intrinsic scale other than $\HVEV$.)
This may seem reminiscent of the argument \cite{Bardeen1995, Lykken2013} 
that the SM, with all masses set to zero, is not FT because it would then acquire classical scale invariance
(broken by quantum $\beta$-functions);
however, our results differ crucially: chiral $SU(2)_{L-R}$
symmetry is actually a quantum symmetry of the Schwinger Model (that is transmitted to the SM), 
while scale invariance is not (and probably isn't for gravity \cite{WeinbergWitten1980}).

To illustrate the absence of  FT in GML
and how that can be preserved when GML is augmented by high-mass-scale physics, 
 we consider, for pedagogical simplicity, an SO(2) version of GML, with a  complex scalar  field $\Phi$,
 with VEV $\HVEV / \sqrt 2$.
In linear and unitary representations, the renormalized field 
$\Phi \equiv\frac{ \left[H+i\sigma_3 \pi_3\right]}{\sqrt{2}}  \equiv \frac {\tilde H}{\sqrt{2}}U; \quad U\equiv \exp\left[{i\sigma_3  {{\tilde \pi}_3}\over{\HVEV} }\right]$
with Pauli matrix $\sigma_3$,
bare  field $\Phi_0 =Z_{\Phi}^{1\over 2} \Phi$, with $\langle H_0 \rangle =Z_{\Phi}^{1\over 2} \HVEV$.
The bare Lagrangian
\begin{eqnarray}
\label{eqn:Lbare-GML}
L^{bare}_{\GMLmth} &=& {1\over2}Tr\vert\partial_\mu\Phi_0\vert^2 - V^{bare}_{\GMLmth;Sym} + L^{bare}_{PCAC} 
\\V^{bare}_{\GMLmth;Sym} &=& {{\mu^2_{\phi (0)}}\over 2} Tr\left[\Phi_0^\dagger\Phi_0\right] + \frac{{\lambda^2_{\phi (0)}}}{4} \left(Tr\left[\Phi_0^\dagger\Phi_0\right]\right)^2  \, .\nonumber
\end{eqnarray}
A Ward-Takahashi Identity (WTI) \cite{Lee1970, Symanzik1970a, Symanzik1970b, Vassiliev1970, ItzyksonZuber} relates bare and renormalized explicit symmetry breaking terms
\begin{equation}
\label{eqn:WardID}
L^{bare}_{PCAC} \equiv \epsilon_0 H_0 = \HVEV m_3^2 H = \epsilon H \equiv L_{PCAC}\,,
 \end{equation}
 with $-m_3^2$ the renormalized $\pi_3$ inverse propagator at $q^2=0$.
 We take the renormalized vacuum, $\HVEV = \langle {\tilde H \rangle}$, 
 to lie in the $\langle {\tilde \pi}_3 \rangle =0$ direction.
 
 Without  $L_{PCAC}$, $L^{bare}_{\GMLmth}$ has an SO(2) chiral symmetry, 
 under which 
 $\Phi \to \exp\left[{i\over 2}\sigma_3\theta\right]\Phi\exp\left[ {i\over 2}\sigma_3\theta\right]$,
easily understood in the unitary representation as a ``shift" symmetry $\tilde \pi_3 \rightarrow \tilde \pi_3 + \HVEV \theta$. 
$L_{PCAC}$ explicitly breaks this chiral SO(2) symmetry,  sourcing only Partial Conservation of the Axial-vector Current (PCAC) $\partial^{\mu}J_{\mu}^5 =\HVEV \mthsq \pi_3$.

In renormalizing $L^{bare}_{\GMLmth}$
we compute quantum loop corrections to its operators 
and replace the 3 bare parameters 
($\mu_{\phi (0)}^2$, $\lambda_{\phi (0)}^2,\epsilon_0$) 
by 3 experimentally measurable quantities $E_i$ $(i\leq 3)$: 
e.g. the quartic coupling $\lambda^2_{\phi}$ and
the renormalized masses-squared $\mhsq , \mthsq$ of the two physical degrees of freedom $h\equiv H - \HVEV$ and $\pi_3$.
Any other measurable quantity $E_j$ ($j \geq 4$) is a function exclusively of $E_i$.    
We call a  theory  ``Physically Fine-Tuned'' if,
for all choices of physical input observables $E_i$,
there exists a (perturbative) physical observable $E_j$ such that for at least one of $E_i$,
${\partial \ln E_j}/{\partial \ln E_i} \gg 1$.
(Certain non-perturbative observables, 
such as the sphaleron-driven rate of baryon-number violation in the SM, are excused because of their exponential dependence on perturbative observables.) 
We make a crucial exception: an observable $E_k$  is not (even bare!) fine-tuned if it is PhysNat,
i.e. if  setting $E_k=0$  increases the symmetry of the theory.      

In \cite{Lynnetal2012}, we examined the renormalization 
with finite Euclidean integral-cutoff $\Lambda$ 
of the UVQD  that appear in the full GML model \cite{GellMannLevy1960} 
(i.e. with a complex scalar doublet,  $SU(2)_L\times SU(2)_R$ broken explicitly to $SU(2)_{L+R}$ by $L_{{\chi}SB}$). 
Renormalizability is transparent only in the linear representation, to which we therefore adhere.
We observed that the UVQD $\sim\Lambda^2$ of the theory were all absorbed into  $\mpisq$, the SU(2) equivalent of $\mthsq$.  
Since setting either $\mpisq$ or $\HVEV$ to zero restores the chiral $SU(2)_{L-R}$ symmetry of the theory, 
GML is PhysNat and neither physically nor bare FT. We extended these results to all perturbative loop-orders.
We observed that including SM fermions (whose Yukawa couplings break the symmetry to $SU(2)_L$) did not alter these conclusions.   (A 4th SM generation fermion
with $m_f^2 \gg \HVEV^2$ {\bf might} be argued FT, but at the expense of in-calculability, with non-perturbative Yukawa coupling $y_f = \sqrt 2 m_{f}/\HVEV \gg 1$, so we ignore that case here)
GML and Schwinger with SM fermions have no BEH-FT problem from UVQDs \cite{Lynnetal2012}.  Neither does the SM \cite{Lynn2011}.  

This conclusion is at odds with the usual  viewpoint, which prefers to cast FT in terms of bare parameters (e.g. $\mu_{\phi(0)}^2$) 
instead of physical ones (e.g. $\mpisq$); 
which prefers the unitary representation of the BEH doublet 
(with which nobody knows how to renormalize)  to the linear one (in which renormalization is straightforward and GML/Schwinger Wigner mode $\HVEV \to 0$ makes sense);
and which prefers to ignore $\mpisq$ rather than treat it as a physical parameter. That view misses the crucial observation: that UVQD are all absorbed into $\mpisq$, and that a zero value for $\mpisq$ is protected by the Goldstone theorem.

In the remainder of this letter, we extend the SO(2) simplification of our GML results \cite{Lynnetal2012} to include certain physics at a new finite scale $M_{Heavy}^2\gg m_h^2$, 
and construct the effective low-energy Lagrangian  $L^{Eff}=L^{bare}+L^{1-loop}$, which emerges after integrating out the heavy degrees of freedom. 
We keep all quadratic ${\cal O}(\Lambda^2)$ and logarithmic ${\cal O}(\ln \Lambda^2)$ divergences, never taking the limit $\Lambda^2 \to \infty$.  
We ignore 5 classes of {\bf finite} operators 
${\cal O}^{Ignore}= 
{\cal O}^{Light}+ {\cal O}^{Heavy}_{marginal}+ {\cal O}^{Heavy}_{constant}+ {\cal O}^{Heavy}_{irrelevant}+{\cal P}^M$: 
${\cal O}^{Light}$ arise entirely from the light degrees of freedom.   
Although important for computing physical observables 
(e.g. the successful 1-loop high precision SM predictions for the top-quark and BEH masses from
Z-pole physics \cite{LynnStuart1985} in 1984
and the $W^{\pm}$ mass \cite{Sirlin1980} in 1980)
they are not the point 
of this letter; 
${\cal O}^{Heavy}_{marginal}$ are marginal operators $\sim \ln(M_{Heavy}^2)$,
e.g. 1st differentials $\Pi_{hh}^{\prime}(q^2), \Pi_{33}^{\prime}(q^2)$ 
of 2-point scalar self-energies evaluated at low $\vert q^2\vert \lapp \mhsq$; 
${\cal O}^{Heavy}_{constant} \sim \left[M_{Heavy}^2 \right]^0$ are analogous with the SM gauge-sector S and U oblique parameters \cite{Kennedy1988,PeskinTakeuchi,Ramond2004} and {\bf might} reveal heavy particles via the scalar sector;
${\cal O}^{Heavy}_{irrelevant}$ are irrelevant operators that vanish as $\mhsq/M_{Heavy}^2\to 0$,
e.g. $\Pi_{hh}^{\prime\prime}(q^2), \Pi_{33}^{\prime\prime}(q^2)$,
2nd and higher differentials with respect to $q^2$
evaluated at low $\vert q^2\vert$; 
${\cal P}^M$ are operators that approach a constant 
 as $q^2$ approaches the physical pole of a 2-point Green's function, 
 and therefore do not contribute to physical observables \cite{Messiah}.
None of these can spoil PhysNat not-FT in GML or the SM..

As shown previously for UVQD \cite{Lynnetal2012}, 
our 1-loop examples will demonstrate explicitly that the WTI (\ref{eqn:WardID}) forces all relevant operator terms --
${\cal O}(\Lambda^2)$, ${\cal O}(M_{Heavy}^2)$, ${\cal O}(\ln \Lambda^2)$, 
${\cal O}(M_{Heavy}^2 \ln M_{Heavy}^2)$ and ${\cal O}(M_{Heavy}^2)$ --
into the renormalized pseudo-NGB mass-squared $\mthsq$, which appears 
with renormalized $\HVEV$ and $\lambda_{\phi}^2$ in the renormalized effective potential $V_{GML}^{Eff}$:
\begin{eqnarray}
\label{eqn:VGML_Ren}
 L^{Eff}_{GML} &=&  {1\over2}Tr\vert\partial_\mu\Phi\vert^2-V_{GML}^{Eff} + {\cal O}^{Ignore} \nonumber \\
V_{GML}^{Eff} &=& \frac{\lambda_\phi^2}{4}\left[{ H}^2 +\pi_3^2 - \left(\HVEV^2 - { {\mthsq} \over{\lambda_\phi^2} } \right) \right]^2  - \HVEV\mthsq {H}  \nonumber
\\&=& {1\over 2} m_h^2 {h}^2 + {1\over 2}\mthsq \pi_3^2 + V_{GML}^{Eff; Cubic, Quartic}
\end{eqnarray}
with $H=h+\HVEV$. A second  WTI \cite{Lee1970} has insisted that
\be
\label{massWTI}
m_h^2=\mthsq + 2 \lambda_\phi^2 \HVEV^2
\ee
at $q^2=0$ in (\ref{eqn:VGML_Ren}). 
The WTI (\ref{eqn:WardID}) ensures the vanishing of the ``tadpole" term (the term linear in $h$) in (\ref{eqn:VGML_Ren}),
automatically
\cite{Lynnetal2012}
enforcing a vacuum stability condition: the BEH cannot simply disappear into the vacuum.   
We see clearly that  SO(2) symmetry is restored in (\ref{eqn:VGML_Ren})
by taking {\bf either} $\mthsq\to0$ or $\HVEV\to0$. Consequently, values of $\mthsq$ or $\HVEV^2$ and thus $m_h^2\ll (\Lambda^2,M_{Heavy}^2)$ are (at least) PhysNat, and neither physically nor bare FT. A crucial observation: {\bf GML/Schwinger Goldstone mode} ($\mthsq \to 0,\HVEV^2 \neq 0,m_h^2 \to  2 \lambda_\phi^2 \HVEV^2$) is also {\bf GoldExceptNat}, with far more powerful suppression of FT.

A so-called FT Problem arises when the term $\HVEV\mthsq H$ is mistakenly ignored while minimizing $V_{GML}^{Eff}$ in (\ref{eqn:VGML_Ren}).
The incorrect result,
$\langle H \rangle _{FT}^2 =\left(\HVEV^2 - { {\mthsq} \over{\lambda_\phi^2} } \right)$, violates GML and Schwinger WTI, violates stationarity \cite{ItzyksonZuber} of the true minimum at $\HVEV$, and destroys the theory's renormalizability and unitarity, which require \cite{Bjorken1965,ItzyksonZuber} wavefunction  renormalization $\HVEV_{Bare}=Z_{\Phi}^{1/2}\HVEV$.

We now examine the consequences of extending SO(2) GML to include a wide class of  high-mass-scale $M_{Heavy}^2\gg m_h^2$ physics.
We display explicit 1-loop results for examples of a heavy scalar and a heavy fermion. 

For the heavy scalar we consider  an SO(2) singlet  real scalar $S$, with  ($S\to-S$) $Z_2$ symmetry, wavefunction renormalization $Z_S$,
$M_S^2\gg m_h^2$, and either $\SVEV =0$ or $\SVEV \neq 0$. 
We add to the renormalized theory
$L_S=\half(\partial_{\mu}S)^2 -V_{\Phi S}$, with 
$V_{\Phi S} = \half\mu_S^2 S^2 + \quarter\lambda_S^2 S^4 + \quarter\lambda_{\phi S}^2 S^2 \left[ Tr(\Phi^\dagger\Phi) -\HVEV^2 \right]$. 

For the heavy fermion we consider an SO(2) singlet right-handed Majorana neutrino $\nu_R$, with $M_{\nu_R}^2\gg m_h^2$, involved in a Type 1 See-Saw with a
left-handed neutrino $\nu_L$, with 
Yukawa coupling $y_{\nu}$ and resulting Dirac mass $m_D=y_{\nu}\HVEV /\sqrt{2}$. 
We add $L_{\nu}=L_{\nu}^{free}+L_{\nu}^{Yukawa}$ to the renormalized theory, with
$L^{free}_{\nu} =i{\bar \nu}_L\partial_{\mu}{\bar \sigma}^{\mu} \nu_L + i{\bar \nu}_R\partial_{\mu}{\bar \sigma}^{\mu} \nu_R - M_{\nu_R} ({\nu_R}{\nu_R}
+{\bar \nu}_R {\bar \nu}_R)/2$,  
$L_{\nu}^{Yukawa}=-y_{\nu} \left[ H(\nu_L \nu_R+{\bar \nu}_L {\bar \nu}_R) -i\pi_3 (\nu_L \nu_R-{\bar \nu}_L {\bar \nu}_R) \right]/{\sqrt 2}$,
${\bar \sigma}^{\mu} = (1,{\vec \sigma})$ with Pauli matrices ${\vec \sigma}$. 

The relevant operator contributions  ${\cal O}(\Lambda^2)$, ${\cal O}(M_{Heavy}^2 \ln \Lambda^2)$, ${\cal O}(M_{Heavy}^2 \ln M_{Heavy}^2)$ and ${\cal O}(M^2_{Heavy})$ to the effective GML Lagrangian of low-mass scalar fields $h$ and $\pi_3$ are examined. We show that, for low-energy $\left|q^2\right| \ll M_{Heavy}^2$ physics, heavy degrees of freedom contribute at worst marginal operators ${\cal O}(\ln M_{Heavy}^2)$.

Tadpole renormalization, as enforced 
by the WTI (\ref{eqn:WardID}), 
relates the renormalized finite  $\mthsq$, the bare mass-squared $m_{3(0)}^2$, and  the $\pi_3$ self-energy at $q^2=0$:
\be
\label{eqn:mthsq}
\mthsq = m_{3(0)}^2 - \Pi_{33}(0)\,.
\ee
Neither $m_{3(0)}^2$ nor $\Pi_{33}(0)$  is a physical observable, 
and  their precise functional forms 
may differ from model to model. Suppressing the contributions of other fermions necessary to make the theory  anomaly-free, we assemble the contributions of $h, \pi_3,\nu_L$ and SO(2) singlets $S, \nu_R$. The crucial calculations appear in the literature \cite{Veltman1981, Farina:2013mla}:
we adapted those for SO(2) with neutrinos and $\SVEV \neq 0$.
\begin{eqnarray}
\label{eqn:BareM3Squared}
&& m_{3(0)}^2=\mu_{\phi (0)}^2 Z_{\phi}+\lambda_{\phi (0)}^2 Z_{\phi}^2 \HVEV^2+{1\over 2}\lambda_{\phi S (0)}^2 Z_{\phi} Z_S \SVEV^2
\\ && 16\pi^2\Pi_{33}(0)= -\lambda_{\phi}^2\left[ 3A(m_h)+A(m_3)\right] -{1\over 2}\lambda_{\phi S}^2 A(M_S) \nonumber
\\ && +y_{\nu}^2 \left[  A(M_{\nu_R}) + A(0)-M_{\nu_R}^2 (\ln{\Lambda^2 \over M_{\nu_R}^2}+1 )-2m_D^2\ln{\Lambda^2 \over M_{\nu_R}^2} \right]      \nonumber
\end{eqnarray}
Because $\Pi_{33}(q^2)$ is UVQD, Passarino-Veltman's \cite{Passarino1979} dimension-2 function
$A(m)=\Lambda^2-m^2\left( \ln{\Lambda^2\over m^2}+1\right)$, for finite Euclidean cutoff $\Lambda$ 
\cite{Veltman1981}, appears. Eqn. (\ref{eqn:BareM3Squared}) then includes all relevant operators: ${\cal O}(\Lambda^2)$, ${\cal O}(M^2_{Heavy}\ln \Lambda^2)$, ${\cal O}(M^2_{Heavy} \ln M^2_{Heavy})$ and ${\cal O}(M^2_{Heavy})$.

It has been falsely claimed that dimensional regularization (DR) eliminates such UVQD. 
We  call this misconception the ``dim-reg herring'', because it sounds good, but is simply untrue.  
As shown by MJG Veltman \cite{Veltman1981}, DR associates UVQD with poles at $d=2$ and 
logarithmic divergences with poles at $d=4$.
Simply throwing them away is no more correct in DR than in any other form of regularization \cite{AlvarezGaume2012,Kennedy1988}.  

Because GML extended
 with high-scale SO(2) representations is renormalizable, 
 the UVQD in $\Pi_{33}(0)$ are exactly cancelled by those in  $m_{3(0)}^2$ \cite{Lynn2011,Lynnetal2012}.
 But what about the $M_S^2$ and $M_{\nu_R}^2$ terms lurking in (\ref{eqn:BareM3Squared})?
If the resulting physical value for $\mthsq$
 is to be ``small,''  i.e. $\ll M_{Heavy}^2$, 
a cancellation must be arranged between $m_{3(0)}^2$and $\Pi_{33}(0)$ in (\ref{eqn:mthsq}).
The smallness of $\mthsq/M_{Heavy}^2$ would normally be called Bare-FT, but setting $\mthsq=0$ in (\ref{eqn:Lbare-GML}) restores chiral SO(2) shift symmetry.  
Therefore $\mthsq/M_{Heavy}^2\ll1$ is PhysNat, and neither physically nor bare FT.     
If $L_{GML}^{bare}$ is extended in  a way that respects that SO(2) symmetry, 
$\mthsq\to0$ remains a symmetry restoration, so is PhysNat not FT.   
The importance of maintaining SO(2) symmetry when extending the model, 
to preserve PhysNat and thus the absence of FT, is a primary lesson of this letter.

We turn next to the properties of the BEH.
In a wide range of $M_{Heavy}^2$ extensions to GML,
$L^{Eff}_{quadratic}$ can readily be shown (using coupled oblique 2-point Dyson's dispersion equations) to be that of free particles, 
when written in terms of the physical fields (${\underbar h},\pi_3$)
and their renormalized  masses-squared ($\mhbsq,m_3^2$).
The underscore on ${\hbbig}$ reflects the possibility that the physical BEH is a linear combination of $h$ 
and one or more other $CP=+1$ scalars.
In GML (and SM), ${\hbbig}=h$.  So too for  a singlet $\nu_R$ with $M_{\nu_R}^2\gg m_h^2$, and for
 an SO(2) singlet scalar $S$  with $M_S^2>>m_h^2$ and $\SVEV=0$. 
In other cases $h$ mixes with a $M_{Heavy}$ field,  e.g. the SO(2) singlet $S$ when $\SVEV \neq 0$.  
${\hbbig}$ is the light field 
(identified with the physical BEH) 
that results from the diagonalization of that mixing, 
and $\mhbsq$ is its pole-mass squared.
$\pi_3$ might also mix with other $CP=-1$ fields, 
but here we confine our attention to theories with no such mixing.  

The pole of the ${\hbbig}$ propagator is given by the relations
\begin{eqnarray}
\label{eqn:mhbsq}
\mhsq &=& \mthsq + 2\lambda_{\phi (0)}^2 Z_\phi^2 \HVEV^2 - \Pi_{\chi}(\mhsq) +  {\cal O}^{Ignore}  \nonumber\\
\mhbsq &=& \mhsq - k_{S\hb} \HVEV^2 
\end{eqnarray}
where $\boldmath{\Pi_{\chi}\equiv\Pi_{hh} - \Pi_{33}}$. 
SO(2) invariance guarantees that, although each of the self-energies $\Pi_{hh}(\mhsq)$ and $\Pi_{33}(\mhsq)$  $\sim \Lambda^2$, 
the combination $\Pi_{\chi} \sim \ln \Lambda^2$, which renormalization group (RG) log divergences are absorbed by the bare Lagrangian term in (\ref{eqn:mhbsq}). 
This matches exactly the absorption, into $\lambda^2_{\phi}$, of the {\bf same} RG divergences $\sim \ln{\Lambda^2}$ in 4-scalar and 3-scalar scattering. 

The term $-k_{S\hb} \HVEV^2$ in (\ref{eqn:mhbsq}) is the expected downward shift in $\mhsq$
when $h$ mixes with the higher mass $S$ to give the physical $\hbbig$.  
In the specific case of our real SO(2) singlet $S$, with $S \to -S$ symmetry and $\SVEV\neq0$,  
$k_{S\hb} = \lambda_{\Phi S}^4 / (2\lambda_S^2)$.
The coupled Dyson dispersion relations show that $ \lambda_{\phi S}^4 (q^2)$ is to be evaluated at 
$q^2=m_{\hb}^2$ 
in determining $k_{S\hb}$. 
Most important is that $-k_{S\hb} \HVEV^2$ is ${\cal O}^{Heavy}_{constant} \sim \left[M^2_{Heavy}\right]^0 \sim {\cal O}^{Ignore}$.
Eqn. (\ref{eqn:mhbsq}) can then be written as in (\ref{eqn:VGML_Ren}).

 What we have found, by explicit 1-loop calculation, for these examples is generic.  Extend GML with SO(2) representations having $M_{Heavy}^2\gg m_h^2$ that respect the chiral SO(2) shift symmetry $\boldmath{{\tilde \pi_3}\to {\tilde \pi_3}+\HVEV \theta}$. 
Then, although each of  $\Pi_{hh}(\mhsq), \Pi_{33}(\mhsq)$ is a relevant operator 
${\cal O}(\Lambda^2)$ or  ${\cal O}(M_{Heavy}^2)$, 
$\Pi_{\chi}(\mhsq)$ in (\ref{eqn:mhbsq}) is at worst ${\cal O}^{Heavy}_{marginal}$. The result can then be written, 
up to ${\cal O}^{Ignore}$, as (\ref{eqn:VGML_Ren}).
 
 It is easy to see that $\mhsq$ is PhysNat and not FT:  (\ref{eqn:mhbsq}) can be re-written
$\mhsq = \mthsq + 2\lambda_{\phi}^2  \HVEV^2$
where $\lambda_\phi^2$ is the renormalized four-point coupling.
Both of these terms are naturally small,
as either $\mthsq\to0$ or $\HVEV\to0$ restores SO(2) symmetry.

The  $SU(2)_L\times SU(2)_R$ Schwinger model \cite{Schwinger1957} is the GML model with 
$L_{\chi SB} \equiv 0$ (i.e. $\epsilon_0 \equiv 0$). 
Over 4 decades, B.W.Lee \cite{Lee1970}, K.Symanzik \cite{Symanzik1970a, Symanzik1970b}, 
and C.Itzkyson and J.B. Zuber \cite{ItzyksonZuber} have emphasized that proper renormalization of GML and Schwinger requires the PCAC term to avoid the BEH Non-Analyticity Problem, namely that 
\be
\label{NonAnalyticity}
\frac{\partial\HVEV^2}{\partial\mu^2}  = -\infty \quad {\rm and} \quad
\frac{\partial\mpisq}{\partial\mu^2} = +\infty
\ee
at the classical scale-invariant point ($\mu^2=m_\pi^2=\HVEV^2 =0$, so that also $\epsilon=\mpisq\HVEV=0$).
(Note the return to the notation $\mpisq$ for the pseudo-NGB mass-squared in the $SU(2)_L\times SU(2)_R$ case.)
{\bf Renormalized Schwinger must be understood as the chiral-symmetric limits,  
$m_{\pi}^2\to 0$ or $\HVEV \to0$, of renormalized GML.} 
This shows the importance of including the PCAC term, 
even in the chiral-symmetric limit.

According to our two examples, the properly renormalized spontaneously broken SO(2) GML/Schwinger low-energy effective Lagrangian is  (\ref{eqn:VGML_Ren}) with $m_3^2 \equiv 0$, 
\begin{eqnarray}
\label{eqn:LSchw_Eff}
 L^{Eff;Goldstone}_{Schwinger}&=& {1\over2}Tr\vert\partial_\mu\Phi\vert^2 - V^{Eff;Goldstone}_{Schwinger} + {\cal O}^{Ignore} \nonumber \\
V^{Eff;Goldstone}_{Schwinger} &=& \lambda_\phi^2\left[ \frac{h^2 +\pi^2_3}{2} + \HVEV h \right]^2
\end{eqnarray}
(\ref{eqn:LSchw_Eff}) includes all ${\cal O}(\Lambda^2),{\cal O}(\ln \Lambda^2)$ UV divergences and all finite relevant operators ${\cal O}^{Heavy}_{relevant} \sim M_{Heavy}^2\gg m_h^2=2\lambda_{\phi}^2 \HVEV^2$, but they have all vanished! GML/Schwinger Goldstone mode therefore suppresses FT far more powerfully than PhysNat, and is instead {\bf GoldExceptNat}, even when extended to include our two heavy-particle examples. 
Explicit 1-loop calculation shows that {marginal operators ${\cal O}^{Heavy}_{marginal} \sim \ln M^2_{Heavy}$ are absorbed by $\lambda_{\phi}^2$ after renormalization}. The heavy fields $S$ and $\nu_R$ {\bf completely decouple} \cite{LSS-1} as $M_S^2$ and $M^2_{\nu_R}\to\infty$ in (\ref{eqn:LSchw_Eff}).

Management of ${\cal O}(\Lambda^2)$, ${\cal O}(M_{Heavy}^2 \ln \Lambda^2)$ and all finite relevant operators ${\cal O}^{Heavy}_{relevant} \sim M_{Heavy}^2$ by the  WTI 
(\ref{eqn:WardID}) has been crucial.  
New (i.e. largely unfamiliar to modern audiences) UVQD \cite{Lynnetal2012} and ${\cal O}(M_{Heavy}^2 \ln \Lambda^2)$ operators, ${\cal O}^{Heavy}_{relevant}$, arise in the 
divergence of the SO(2) GML ($\epsilon_0 \neq 0$) axial-vector current $\partial^{\mu}J_{\mu}^5$, 
which do not arise in Schwinger ($\epsilon_0 =0$) axial-vector current conservation. 
However, WTI 
(\ref{eqn:WardID}) 
ensures that these new ${\cal O}(\Lambda^2)$, ${\cal O}(M_{Heavy}^2 \ln \Lambda^2)$ and ${\cal O}^{Heavy}_{relevant}$ terms, 
together with those in $\mthsq$ in 
(\ref{eqn:mthsq}),
obey the PCAC relation $\partial^{\mu}J_{\mu}^5=\HVEV\mthsq \pi_3$, and collectively vanish as 
$\mthsq\to0$. 
Eq. (\ref{eqn:WardID}) is the first of a tower of recursive WTI governing connected amputated one-particle-irreducible 
Greens functions \cite{Lee1970, LSS-1}.
These guarantee that SO(2) symmetry is restored exactly in the limits $\mthsq\to0$ {\bf and/or} $\HVEV \to 0$. 

 Vast experimental and observational evidence shows that the SM must be extended to include at least classical General Relativistic (GR) gravity.  
 The appearance of  quantum gravity (QG)  by/at the Planck Scale $M_{Planck}$ 
 is often given as the most fundamental reason for worrying about the BEH-FT problem,
 and the reason why showing there is no BEH-FT problem in
 the pure SM ``is not good enough".   
 We reiterate our previous statement \cite{Lynnetal2012}: if the SM were somehow shown \cite{LSS-2} to be GoldExceptNat not FT, then preserving that 
is a problem for those seeking to extend the SM,
 not for the SM.

In this letter we have shown that, {\bf contrary to common wisdom}, 
SO(2) GML and Schwinger satisfy G. 't Hooft's physical naturalness no-FT criteria, while GML/Schwinger Goldstone mode satisfies the criteria for GoldExceptNat, with its far more powerful suppression of FT. This remains true for certain heavy-particle extensions: e.g. singlet $\nu_R$ with large Majorana masses
and additional massive singlet scalars with large VEVs. 

If Goldstone Exceptional Naturalness could somehow be extended to the SM \cite{LSS-2}, phenomenological consequences of the absence of SM BEH FT would include: 
invalidating an upper BEH-FT bound 
on $M_{\nu_R}^2$ from (\ref{eqn:BareM3Squared}), 
thereby eliminating an argument  \cite{Espinosa,Farina:2013mla} against Type 1 See-Saw early universe thermal lepto-genesis \cite{Yanagida1986} 
and subsequent baryo-genesis 
\cite{Rubakov1985,Rubakov1998} 
in the $\nu$MSM 
\cite{Shaposhnikov2005a,Shaposhnikov2005b};
invalidating an upper BEH-FT bound from (\ref{eqn:BareM3Squared})  on the singlet scalar mass $M_S$, which had placed it within range of LHC \cite{Farina:2013mla};
most importantly, 
there should be no {\bf expectation} \cite{Lynn2011} that LHC will discover 
BSM physics unrelated to neutrino mixing, the only known experimentally necessary modification of the SM+GR paradigm.

{\bf Acknowledgments:}
We thank A. Matas  and  R. Stora for illuminating conversations,  
and L. Alvarez-Gaume and J. Donoghue for conversations regarding quantum gravity's
affect on these results.
BWL thanks Jon Butterworth and the Physics/Astronomy Dept. at University College London for support. 
 GDS is partially supported by CWRU grant DOE-SC0009946,   
and thanks the CERN theory group for hospitality during 2012-2013.

 \end{document}